\documentclass[10pt,osajnl2,preprint,showpacs,superscriptaddress,twocolumn]{revtex4}  

\usepackage{hyperref} 

\usepackage{graphicx}
\usepackage{subfigure}
\usepackage{amsmath, amsthm, amssymb}
\usepackage{color}

\renewcommand\Re{\operatorname{Re}}
\renewcommand\Im{\operatorname{Im}}

\begin{document}

\title{Modeling dielectric half-wave plates for CMB polarimetry using a Mueller matrix formalism}
\author{Sean A. Bryan}
 \affiliation{CERCA, Department of Physics, Case Western Reserve University, \\ 10900 Euclid Avenue, Cleveland, OH, 44106, USA}
 \email{sean.bryan@case.edu}
\author{Thomas E. Montroy}
 \affiliation{CERCA, Department of Physics, Case Western Reserve University, \\ 10900 Euclid Avenue, Cleveland, OH, 44106, USA}
\author{John E. Ruhl}
 \affiliation{CERCA, Department of Physics, Case Western Reserve University, \\ 10900 Euclid Avenue, Cleveland, OH, 44106, USA}
 
 
 
\begin{abstract}
We derive an analytic formula using the Mueller matrix formalism that parameterizes the non-idealities of a half-wave plate (HWP) made from dielectric AR-coated birefringent slabs. This model accounts for frequency-dependent effects at normal incidence, including effects driven by the reflections at dielectric boundaries.
The model also may be used to guide the characterization of an instrument that uses a HWP.  We discuss the coupling of a HWP to different source spectra, and the potential impact of that effect on foreground removal for the SPIDER CMB experiment.  We also describe a way to use this model in a map-making algorithm that fully corrects for HWP non-idealities.
\end{abstract}

\maketitle

\section{Introduction}
CMB polarization encodes information about Inflation~\cite{baumann09}, 
re-ionization~\cite{zaldarriaga08}, and the
large-scale structure of the universe \cite{smith08}. 
Many upcoming
experiments including SPIDER \cite{crill08}, EBEX \cite{sagiv10},
POLARBEAR \cite{lee08}, Keck \cite{sheehy10}, ABS \cite{essinger-hileman09} and others will
use a half-wave plate (HWP) to modulate the polarization state of
light from the sky before measurement by the detectors.
Such HWPs may be periodically stepped (eg between maps of the same part of the sky)
to reduce the effect of beam asymmetries and instrumental polarization of optical elements skyward of the HWP.  Alternatively, 
they can be continuously rotated to modulate the signal and to also reject atmospheric variations and 1/f noise~\cite{johnson06}.

An ideal HWP with one of its crystal axes oriented at an angle $\theta_{hwp}$ to the plane of polarization of incident light rotates that polarization plane by  $2\theta_{hwp}$ as the light passes through it.  Real HWPs made from birefringent materials have several important non-idealities. Since the phase delay is only a half-wave at a single frequency, the exact angle through which a HWP rotates the polarization state is frequency-dependent.  Additionally, reflections from the material interfaces reduce transmission and induce non-ideal rotation;  to minimize these effects anti-reflection (AR) coatings are typically used on the surfaces of the HWP.  However,  AR coatings are frequency dependent and do not fully eliminate these non-ideal behaviors.  

These effects have been treated in a variety of ways by other authors.  O'Dea et. al. \cite{odea07} and Brown et. al. \cite{brown09} used a Mueller matrix formalism to parameterize a HWP and polarized detector, but did not connect their parameterization with a physical model of a HWP. Savini et. al. \cite{savini06} described in detail a physical model for stacks of dielectric birefringent materials. Their model works directly with the electric fields, which allows it to handle the input and output polarization state of the light, multiple reflections from dielectric interfaces, and the finite bandwidth over which the stack is a half-wave retarder. 
Matsumura \cite{matsumura06} modeled a multiple-layer HWP using a similar approach.  We employ the methods of Savini et. al. \cite{savini06} in this paper, using Jones and Mueller matrix methods to derive exact couplings from sources (of arbitrary but known spectra) on the sky to a detector, through a non-ideal HWP made of any number of dielectric layers.  Our resulting model can be evaluated quickly in observing simulations such as those done for SPIDER~\cite{mactavish08}, an important property for upcoming CMB experiments.  Additionally, we show how to use this model to correct for known HWP non-idealities during the mapmaking process (using an input data timestream to make a map of intensity and polarization on the sky), greatly reducing the potential impact of systematic effects induced by such non-idealities.

\section{The HWP Mueller Matrix}
Our goal is to calculate the combined Mueller matrix of a HWP plus polarized-detector system for arbitrary orientations of the instrument and HWP relative to the coordinates defining the Stokes parameters of the incoming radiation, including the effects of reflections at the various dielectric interfaces and the effect of band averaging.  A partially polarized detector can be modeled as a partial linear polarizer with Mueller matrix $\mathsf{M}_{pol}$ followed by a total power detector that is sensitive only to the $I$ Stokes parameter.  The total Mueller matrix $\mathsf{M}$ is found by combining the Mueller matrices of the partial polarizer and the HWP, taking rotational alignments into account.  The total detected power $d$ for incident light with a Stokes vector $S = [I, Q, U, V]$ is given by summing over the top row the product of $\mathsf{M}$ and $S$~\cite{jones08},
\begin{equation} \label{eqn_one}
d = I  M_{II} + Q  M_{IQ} + U  M_{IU} + V  M_{IV} ,
\end{equation}
where the $M_{XX}$ are elements of the top row of the total Mueller matrix.

Since physical models of the action of a birefringent HWP are constructed using electric fields, we start the calculation in the Jones formalism. A Jones matrix is a 2-by-2 matrix of complex numbers that that describes the action of an optical system on the $x$- and $y$-components of the electric field of an incident plane wave. 


We start by considering the Jones matrix of a general retarder,
\begin{equation}\label{j_hwp}
\mathsf{J}_{ret}(f) = \left[ \begin{array}{cc} a(f) & \epsilon_1(f)  \\ \epsilon_2(f) & b(f) e^{i \phi(f)} \end{array} \right],
\end{equation}
where 
$a(f)$, $b(f)$, and $\phi(f)$ are real, and $\epsilon_1(f)$ and $\epsilon_2(f)$ are small and complex~\cite{odea07}. 
To convert this to a Mueller matrix, we follow Jones et. al.~\cite{jones08} and use
\begin{equation}\label{jones2mueller}
M_{ij} = \frac{1}{2} \mathrm{trace} ( \sigma_i \mathsf{J} \sigma_j \mathsf{J}^\dagger )
\end{equation}
from Born and Wolf \cite{born}, where $\sigma_i$ are the Pauli matrices,
\begin{eqnarray}
\begin{array}{cc}
\sigma_1 = \left[ \begin{array}{cc} 1 & 0  \\ 0 & 1 \end{array} \right] &
\sigma_2 = \left[ \begin{array}{cc} 1 & 0  \\ 0 & -1 \end{array} \right] \\
 & \\
\sigma_3 = \left[ \begin{array}{cc} 0 & 1  \\ 1 & 0 \end{array} \right] &
\sigma_4 = \left[ \begin{array}{cc} 0 & -i  \\ i & 0 \end{array} \right]. \\
\end{array}
\end{eqnarray}
This will yield the Mueller matrix as a function of frequency $\mathsf{M}_{ret}(f)$ of the HWP. Since Mueller matrices can be band-averaged, we integrate this single-frequency Mueller matrix against a CMB or foreground spectrum $S(f)$, as well as the detector passband $F(f)$. This gives the band-averaged Mueller matrix
\begin{equation}\label{band_average}
\mathsf{M}_{HWP} = \frac{\int df \mathsf{M}_{ret}(f)  S(f) F(f)}{\int df S(f) F(f)}.
\end{equation}

\subsection{Single-Plate HWP}

For a HWP made from a single layer of birefringent dielectric material, $x$- and $y$-polarized states defined in the crystal axes cannot couple into each other. This means that in the HWP Jones matrix of Eq.~\ref{j_hwp}, $\epsilon_1(f)~=~\epsilon_2(f)~=~0$. This leaves only three parameters $a(f)$, $b(f)$ and $\phi(f)$ that are necessary to completely characterize the HWP. The transmission coefficients $a(f)$ and $b(f)$ vary with frequency because of the passband of the AR coating and the interference of multiple reflections inside the birefringent layer. The relative phase delay $\phi(f)$ also varies with frequency because the path length difference for polarization states traveling along the slow and fast crystal axes is $(n_s - n_f) d = \frac{\phi(f)}{2 \pi} \frac{c}{f}  $.

Applying Eq.~\ref{jones2mueller} to the Jones matrix of a single-plate HWP
\begin{equation}\label{j_single_plate_hwp}
\mathsf{J}_{ret}(f) = \left[ \begin{array}{cc} a(f) & 0  \\ 0 & b(f) e^{i \phi(f)} \end{array} \right]
\end{equation}
gives the corresponding Mueller matrix $\mathsf{M}_{ret}(f)$
\begin{eqnarray}\label{m_ret}
\left[ \begin{array}{cccc} \frac{1}{2}(a^2 + b^2) & \frac{1}{2}(a^2 - b^2) & 0 & 0 \\ \frac{1}{2}(a^2 - b^2) & \frac{1}{2}(a^2 + b^2) & 0 & 0  \\ 0 & 0 & a b \cos(\phi) & - a b \sin(\phi) \\ 0 & 0 & a b \sin(\phi) & a b \cos(\phi) \end{array} \right],
\end{eqnarray}
where $a$, $b$, and $\phi$ are all functions of frequency. This reduces to the result given in Tinbergen \cite{tinbergen96} for an ideal retarder ($a = b = 1$). 

\begin{figure*}
\begin{center}
\includegraphics[width=1.0\textwidth]{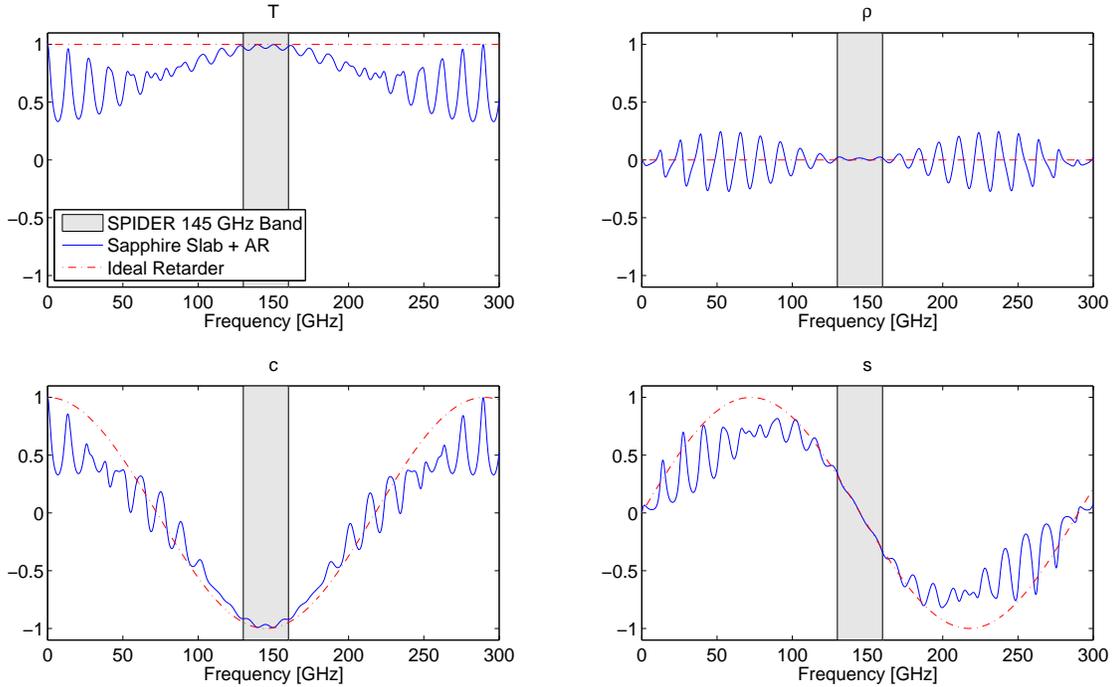}
\caption{Calculated Mueller matrix elements (see Eq.~\ref{m_hwp} for where these elements lie in the matrix) as a function of frequency of an optimized SPIDER 145 GHz HWP (solid blue lines). The dot-dash red lines show the Mueller matrix elements of a HWP retarder with a perfect AR coating for comparison. The grey box shows the SPIDER 145 GHz band. \label{hwp_mueller}}
\end{center}
\clearpage 
\end{figure*}

In general, the band-averaged HWP Mueller matrix $\mathsf{M}_{HWP}$ calculated using Eq.~\ref{band_average} will not be of the same form as $\mathsf{M}_{ret}(f)$ and will have four (rather than three) independent non-zero elements,
\begin{equation}\label{m_hwp}
\mathsf{M}_{HWP} \equiv \left[ \begin{array}{cccc} T & \rho & 0 & 0 \\ \rho & T & 0 & 0  \\ 0 & 0 & c & -s \\ 0 & 0 & s & c \end{array} \right].
\end{equation}
For the specific case of the SPIDER 145 GHz HWP, we have calculated this Mueller matrix as a function of frequency, and plotted it in Fig.~\ref{hwp_mueller}. Since the CMB is not expected to be circularly-polarized, in the case where the subsequent detector and optical system does not induce sensitivity to circular polarization the $s$ parameter will not be relevant for CMB polarimetry.

\subsection{Multiple-Plate HWP}

For a multiple-layer HWP, we cannot simply multiply together the Mueller matrices of several single-layer HWPs, because this would not account for multiple reflections among the interfaces between each layer. We instead return to the original Jones matrix of a general retarder in Eq.~\ref{j_hwp}. For a multiple-layer HWP, $\epsilon_1$ and $\epsilon_2$ will be small, but still non-zero. This means that all seven of the parameters of the Jones matrix, $a(f)$, $b(f)$, $\phi(f)$, $\Re[\epsilon_1(f)]$, $\Im[\epsilon_1(f)]$, $\Re[\epsilon_2(f)]$, and $\Im[\epsilon_2(f)]$ are required to characterize the HWP. After applying Eq.~\ref{jones2mueller}, we obtain the corresponding Mueller matrix
\begin{widetext}
\begin{eqnarray}
\mathsf{M}_{ret}(f) = \left[ \begin{array}{cccc}
\frac{1}{2} \left( a^{2} + b^{2} + |\epsilon_{1}|^{2} + |\epsilon_{2}|^{2} \right) & \frac{1}{2} \left( a^{2} - b^{2} - |\epsilon_{1}|^{2} + |\epsilon_{2}|^{2} \right) & \dots \\
\frac{1}{2} \left( a^{2} - b^{2} + |\epsilon_{1}|^{2} - |\epsilon_{2}|^{2} \right) & \frac{1}{2} \left( a^{2} + b^{2} - |\epsilon_{1}|^{2} - |\epsilon_{2}|^{2} \right) & \dots \\
a \cdot \Re[\epsilon_{2}] + b \cdot \left( \Re[\epsilon_{1}] \cos(\phi) + \Im[\epsilon_{1}] \sin(\phi) \right) & a \cdot \Re[\epsilon_{2}] - b \cdot \left( \Re[\epsilon_{1}] \cos(\phi) + \Im[\epsilon_{1}] \sin(\phi) \right) & \dots \\
a \cdot \Im[\epsilon_{2}] + b \cdot \left( \Re[\epsilon_{1}] \sin(\phi) - \Im[\epsilon_{1}] \cos(\phi) \right) & a \cdot \Im[\epsilon_{2}] - b \cdot \left( \Re[\epsilon_{1}] \sin(\phi) - \Im[\epsilon_{1}] \cos(\phi) \right) & \dots
\end{array} \right. \nonumber \\
\left. \begin{array}{cccc}
\textrm{Row~1~cont.} & a \cdot \Re[\epsilon_{1}] + b \cdot \left( \Re[\epsilon_{2}] \cos(\phi) + \Im[\epsilon_{2}] \sin(\phi) \right) & - a \cdot \Im[\epsilon_{1}] - b \cdot \left( \Re[\epsilon_{2}] \sin(\phi) - \Im[\epsilon_{2}] \cos(\phi) \right) \\
\textrm{Row~2~cont.} & a \cdot \Re[\epsilon_{1}] - b \cdot \left( \Re[\epsilon_{2}] \cos(\phi) + \Im[\epsilon_{2}] \sin(\phi) \right) & - a \cdot \Im[\epsilon_{1}] + b \cdot \left( \Re[\epsilon_{2}] \sin(\phi) - \Im[\epsilon_{2}] \cos(\phi) \right)  \\
\textrm{Row~3~cont.} & \Re[\epsilon_{1}] \cdot \Re[\epsilon_{2}] + \Im[\epsilon_{1}] \cdot \Im[\epsilon_{2}] + a b \cos(\phi)& \textcolor{white}{-}\Re[\epsilon_{1}] \cdot \Im[\epsilon_{2}] - \Im[\epsilon_{1}] \cdot \Re[\epsilon_{2}] - a b \sin(\phi) \\
\textrm{Row~4~cont.} & \Re[\epsilon_{1}] \cdot \Im[\epsilon_{2}] - \Im[\epsilon_{1}] \cdot \Re[\epsilon_{2}] + a b \sin(\phi) & -\Re[\epsilon_{1}] \cdot \Re[\epsilon_{2}] - \Im[\epsilon_{1}] \cdot \Im[\epsilon_{2}] + a b \cos(\phi) 
\end{array} \right].
\end{eqnarray}
\end{widetext}
This reduces to the single plate result of Eq.~\ref{m_ret} if $\epsilon_{1} = \epsilon_{2} = 0$. Note that none of the elements of this matrix have the same functional form. This means that after applying Eq.~\ref{band_average} to band-average against the source and detector spectra, all 16 elements of the resulting Mueller matrix will be independent. Since the CMB is not expected to be circularly-polarized and the subsequent optical and detector system should have no sensitivity to circular polarization, only the top-left 9 elements of the Mueller matrix will be relevant for CMB polarimetry.

\section{Rotating the Instrument and HWP}

Jones et. al.~\cite{jones08} modeled a polarization-sensitive detector as a rotatable instrument with a partial-polarizer followed by a total power detector.   The Jones matrix of a vertical partial polarizer is
\begin{equation}\label{j_pol_a}
\mathsf{J}_{pol} = \left[ \begin{array}{cc} \eta & 0  \\ 0 & \delta \end{array} \right],
\end{equation}
which can be turned into a corresponding Mueller matrix
\begin{eqnarray}\label{m_pol}
\mathsf{M}_{pol} = \left[ \begin{array}{cccc} \frac{1}{2}(\eta^2 + \delta^2) & \frac{1}{2}(\eta^2 - \delta^2) & 0 & 0 \\ \frac{1}{2}(\eta^2 - \delta^2) & \frac{1}{2}(\eta^2 + \delta^2) & 0 & 0  \\ 0 & 0 & \eta \delta & 0 \\ 0 & 0 & 0 & \eta \delta \end{array} \right].
\end{eqnarray}
Given the instrument rotation matrix~\cite{tinbergen96}
\begin{equation}
\mathsf{M}_{\psi} = \left[ \begin{array}{cccc} 1 & 0 & 0 & 0 \\ 0 & \textcolor{white}{-}\cos{2 \psi_{inst}} & \sin{2 \psi_{inst}} & 0  \\ 0 & -\sin{2 \psi_{inst}} & \cos{2 \psi_{inst}} & 0 \\ 0 & 0 & 0 & 1 \end{array} \right],
\end{equation}
the detected radiation is given by the coupling to $I$ in the product $\mathsf{M}_{pol}~\mathsf{M}_{\psi}$;  for an 
ideal polarizer with $\eta = 1$ and $\delta = 0$ and an arbitrary instrument angle $\psi_{inst}$, the detector signal is
\begin{equation}
d = \frac{1}{2} \left(I + Q \cos(2 \psi_{inst}) + U \sin(2 \psi_{inst}) \right).
\end{equation}

\begin{figure}
\begin{center}
\includegraphics[width=0.48\textwidth]{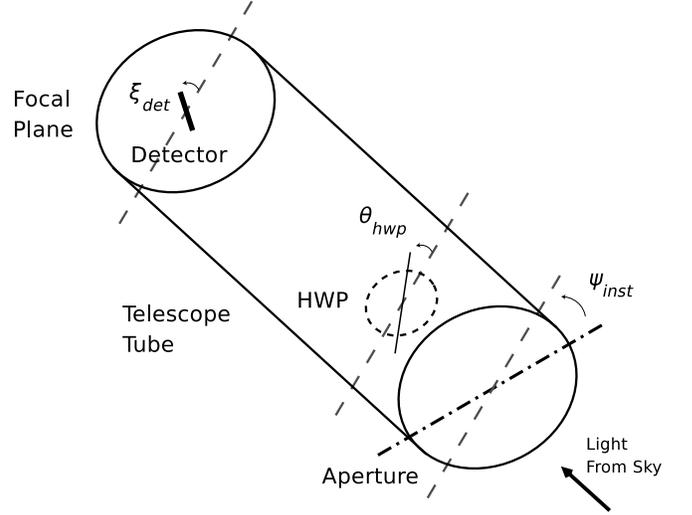}
\caption{Our definitions of the detector, HWP, and instrument angles. The instrument angle is defined relative to a fixed reference on the sky that determines the absolute orientations of $Q$ and $U$, and the detector and HWP angles are defined relative to the instrument. \label{instrument_and_hwp_angles}}
\end{center}
\clearpage 
\end{figure}

The addition of a HWP to the instrument can be modeled with the Mueller matrix product
\begin{equation}\label{mat_multiply}
\mathsf{M} = \mathsf{M}_{pol}~\mathsf{M}_\xi~\mathsf{M}_{-\theta}~\mathsf{M}_{HWP}~\mathsf{M}_{\theta}~\mathsf{M}_{\psi},
\end{equation}
where $\mathsf{M}_{\theta}$ is the rotation matrix by the HWP angle $\theta_{hwp}$, and $\mathsf{M}_{\xi}$ is the rotation matrix by the detector orientation angle $\xi_{det}$. 
Fig.~\ref{instrument_and_hwp_angles} illustrates the definition of these angles;  while only two
angles are needed to define the single-detector problem, we use three here to make the problem more straightforward to visualize, and to easily accommodate calculations of focal planes with 
detectors at multiple orientation angles.  Here we consider only a single detector with 
$\xi_{det} = 0$.

The top row of the resulting Mueller matrix for a single-plate HWP is
\begin{eqnarray}\label{hwp_matrix_elements}
M_{II} &=& \frac{1}{2} \left[ T (\eta^2 + \delta^2) + \rho \cos(2 \theta_{hwp}) (\eta^2 - \delta^2) \right] \nonumber \label{start_of_big_equation} \\
M_{IQ} &=&  \textcolor{white}{-}\mathcal{F} \sin(2 \psi_{inst})  +  \mathcal{G} \cos(2 \psi_{inst}) \nonumber \\
M_{IU} &=&  - \mathcal{F} \cos(2 \psi_{inst})  + \mathcal{G} \sin(2 \psi_{inst}) \nonumber \\
M_{IV} &=& \frac{1}{2} s \sin(2 \theta_{hwp}) (\eta^2 - \delta^2),
\end{eqnarray}
where $\mathcal{F}$ and $\mathcal{G}$ are
\begin{eqnarray}
\mathcal{F} &\equiv&   - \frac{1}{4} (T-c) \sin(4 \theta_{hwp}) (\eta^2 - \delta^2) \nonumber \\
&-& \frac{1}{2} \rho \sin(2 \theta_{hwp})(\eta^2 + \delta^2) \\
\mathcal{G} &\equiv& \frac{1}{4} \left[T + c + (T-c) \cos(4 \theta_{hwp}) \right] (\eta^2 - \delta^2) \nonumber \\
&+& \frac{1}{2} \rho \cos(2 \theta_{hwp}) (\eta^2 + \delta^2). \label{end_of_big_equation}
\end{eqnarray}
This reduces to the result given in Jones et. al. \cite{jones08} in the limit of no HWP ($T = c = 1,~\rho = s = 0$), and in the case of an ideal HWP ($T=1,~c=-1,~\rho = s = 0$) reduces to
\begin{eqnarray}
M^{ideal}_{II} &=& \frac{1}{2} (\eta^2 + \delta^2) \nonumber \\
M^{ideal}_{IQ} &=&   \frac{1}{2} \cos(2 ( \psi_{inst} + 2 \theta_{hwp})) (\eta^2 - \delta^2) \nonumber \\
M^{ideal}_{IU} &=&  \frac{1}{2} \sin(2 ( \psi_{inst} + 2 \theta_{hwp})) (\eta^2 - \delta^2) \nonumber \\
M^{ideal}_{IV} &=& 0.
\end{eqnarray}

Armed with the band-averaged Mueller matrix representations of the HWP plus detector system given by Eq.~\ref{mat_multiply}, we can use 
Eq.~\ref{eqn_one} to calculate the detector output as a function of input Stokes parameters.


\begin{figure*}[htb]
\begin{center}
\subfigure{\includegraphics[width=0.49\textwidth]{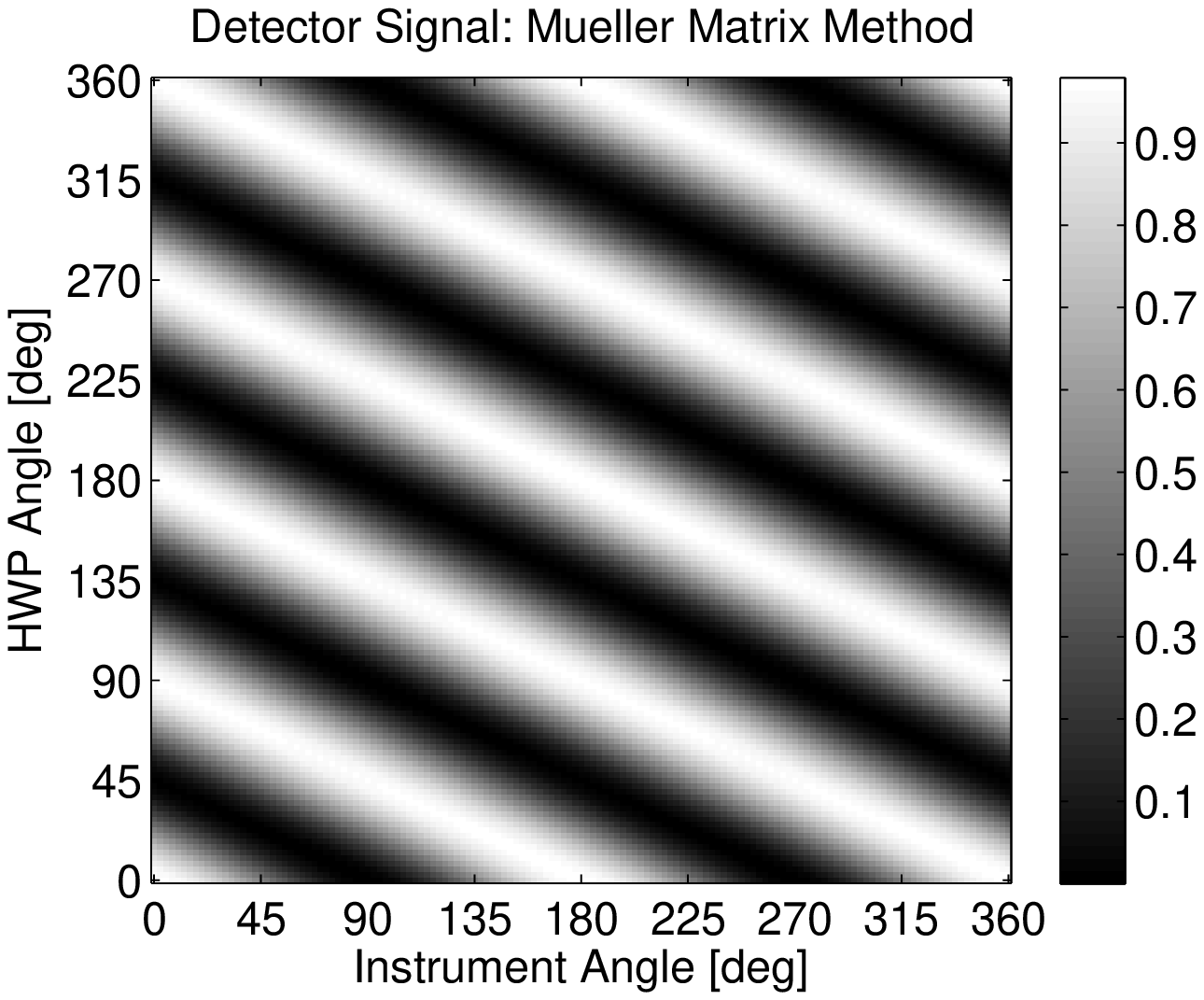}} 
\subfigure{\includegraphics[width=0.49\textwidth]{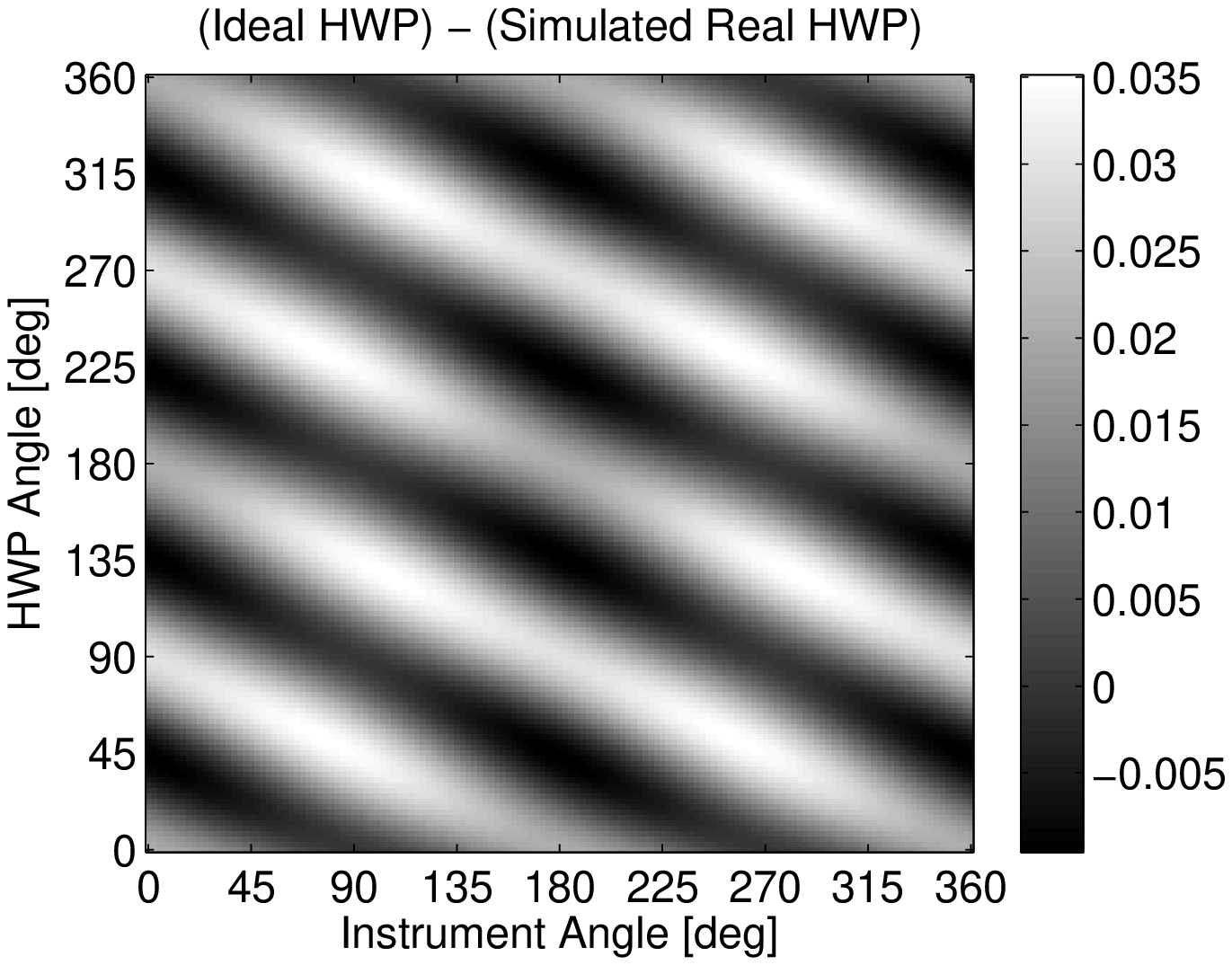}} 
\caption{Simulated detector output for the SPIDER instrument as a function of HWP and instrument angle for a $Q$-polarized CMB $dB/dT$ source. The left panel shows the detector output calculated using our band-averaged Mueller matrix formalism.
The right panel shows the percent-level differences between an ideal HWP and the simulated non-ideal HWP. 
\label{cmb_sim}}
\end{center}
\clearpage 
\end{figure*}

An example of the output from our model, along with the percent-level differences between that and a naiive treatment, is shown in Fig.~\ref{cmb_sim} for the specific case of the SPIDER experiment, which uses a single birefringent sapphire HWP with quarter-wave AR coats.  

\section{Detector Pair Summing and Differencing}

Many CMB polarization experiments use pairs of Polarization-Sensitive Bolometers (PSBs) located at the same point on the focal plane \cite{jones02}. Both detectors in a pair view the same patch of the sky, but detector B is oriented at $90^\circ$ with respect to detector A. Detector differencing within a pair reduces common-mode noise while retaining sensitivity 
to linear polarization. The sum of a pair measures the $I$ Stokes parameter. 
We take Eq.~\ref{eqn_one} with $\mathsf{M}$ given by Eq.~\ref{mat_multiply} with $\xi_{det} = 0$ as a model for the A detector timestream $d_i^A$. The B detector timestream $d_i^B$ can be similarly calculated by setting
$\xi_{det} = 90^\circ$;  we note however that the Jones matrix of a horizontal polarizer,
\begin{equation}\label{j_pol_b}
\mathsf{J}_{pol}^B = \left[ \begin{array}{cc} \delta & 0  \\ 0 & \eta \end{array} \right],
\end{equation}
is related to that of a vertical polarizer (Eq.~\ref{j_pol_a}) via the substitutions
$\eta \rightarrow \delta$ and $\delta \rightarrow \eta$.  We can thus just make those substitutions in Equations \ref{start_of_big_equation} through \ref{end_of_big_equation} to find a model for the B detector timestream. We then construct the sum and difference timestreams
\begin{eqnarray}
d_i^{sum} &\equiv&  d^A_i + d^B_i , \nonumber\\
d_i^{diff} &\equiv&  d^A_i - d^B_i.
\end{eqnarray}
Only  $(\delta^2 + \eta^2)$ terms will remain in the sum timestream and only $(\delta^2 - \eta^2)$ terms will remain in the difference timestream. The matrix elements for the sum timestream with a single-plate HWP are therefore
\begin{eqnarray}
M_{II}^{sum} &=& T (\eta^2 + \delta^2) \nonumber \\
M_{IQ}^{sum}  &=&  \rho \cos(2 (\psi_{inst} + \theta_{hwp}) ) (\eta^2 + \delta^2) \nonumber \\
M_{IU}^{sum}  &=&  \rho \sin(2 (\psi_{inst} + \theta_{hwp}) ) (\eta^2 + \delta^2) \nonumber \\
M_{IV}^{sum}  &=& 0.
\end{eqnarray}
Even with the non-idealities of the HWP, the sum timestream coupling to intensity is independent of HWP angle. There is also a small coupling to linear polarization. The matrix elements for the difference timestream are
\begin{eqnarray}
M_{II}^{diff} &=& \rho \cos(2 \theta_{hwp}) (\eta^2 - \delta^2) \nonumber \\
M_{IQ}^{diff}  &=&  \textcolor{white}{-}\left[ \mathcal{F}'   \sin(2 \psi_{inst}) + \mathcal{G}'  \cos(2 \psi_{inst}) \right] (\eta^2 - \delta^2) \nonumber \\
M_{IU}^{diff}  &=&  \left[-\mathcal{F}' \cos(2 \psi_{inst}) + \mathcal{G}'\sin(2 \psi_{inst}) \right] (\eta^2 - \delta^2) \nonumber \\
M_{IV}^{diff}  &=& s \sin(2 \theta_{hwp}) (\eta^2 - \delta^2) ,
\end{eqnarray}
where $\mathcal{F}'$ and $\mathcal{G}'$ are
\begin{eqnarray}
\mathcal{F}' &\equiv&    -\frac{1}{2} (T-c) \sin(4 \theta_{hwp}) \\
\mathcal{G}' &\equiv& \frac{1}{2} \left[T + c + (T-c) \cos(4 \theta_{hwp}) \right] .
\end{eqnarray}
The difference timestream unfortunately has small couplings to intensity and circular polarization. Note that for both the sum and difference timestreams, detector cross-polarization only shows up as an overall factor of $(\eta^2~\pm~\delta^2)$, and will not result in leakage between the estimates of $Q$ and $U$.

\section{Application to a Specific Instrument}

The model we derived requires input values for the HWP parameters $a(f)$, $b(f)$, $\phi(f)$, $\epsilon_{1}(f)$, and $\epsilon_{2}(f)$.  We can calculate these parameters from first principles, or measure them using a polarized Fourier transform spectrometer (FTS) with the assembled instrument.

\subsection{Physical Modeling}
To calculate the HWP parameters from first principles, we use a physical optics model similar to the one in Savini et al. \cite{savini06}. The model extends the 2-by-2 matrix formalism described by Hecht and Zajac~\cite{hecht74} for modeling multiple dielectric layers of isotropic materials to a 4-by-4 matrix formalism for multiple layers of potentially birefringent materials. The model uses the electromagnetic boundary conditions at the interface between each layer of material to map the incident electric and magnetic fields onto the transmitted fields. This fully treats multiple reflections and interference effects, and can also handle lossy materials.

We use this model to calculate the elements of the HWP Jones matrix shown in Eq.~\ref{j_hwp} by
calculating the transmitted electric field amplitude $\vec{E}_{out}$
at a frequency $f$ with an incident electric field amplitude $\vec{E}_{in}$, for both
$x$- and $y$-polarized incident waves.
We assemble the transmitted amplitudes into the Jones matrix for that HWP,
\begin{eqnarray}\label{hwp_jones_calculation}
\left[ \begin{array}{c} a(f) \\ \epsilon_2(f) \end{array} \right] = \vec{E}_{out} \left(\left[ \begin{array}{c} 1 \\ 0 \end{array} \right],f \right), \nonumber \\
\left[ \begin{array}{c} \epsilon_1(f) \\ b(f) e^{i \phi(f)} \end{array} \right] = \vec{E}_{out} \left(\left[ \begin{array}{c} 0 \\ 1 \end{array} \right],f \right).
\end{eqnarray}


\subsection{Instrument Calibration}

The parameterization we developed can also be used to characterize the HWP, with the aim of better understanding the coupling to intensity and linear polarization signals from the CMB and foregrounds with other spectra.  This characterization process depends on the manner in which the HWP will be used, ie whether it will be continuously rotated or occasionally stepped during the observation process. For both the stepped and continuously rotating cases, a polarized FTS can be used to measure the frequency and polarization response of the instrument and HWP. Taking the Fourier transform of the detector timestream as the FTS mirror moves at a constant speed will yield the spectral response of the combined instrument \cite{lesurf90}.

\subsubsection{Stepped HWP}

As an illustration of how the calibration process of an instrument with a stepped 
single-plate 
HWP might work, consider the case where we align the instrument with the FTS such that 
$\xi_{det}=\psi_{inst} = 0^\circ$. The FTS spectrum $\mathcal{S}(\theta_{hwp},f)$ at a particular HWP angle can be modeled using 
Eq.~\ref{eqn_one} with the Mueller matrix elements given in Eq.~\ref{mat_multiply}. Before proceeding, it is useful to change variables from $\eta$ and $\delta$ to the optical efficiency and the polarization efficiency $\gamma$. Here we define the optical efficiency as $\varepsilon \equiv \frac{1}{2}(\eta^2 + \delta^2)$ and follow Jones et. al. \cite{jones08} and take the polarization efficiency $\gamma \equiv \frac{\eta^2 - \delta^2}{\eta^2 + \delta^2}$. Setting $\psi_{inst} = 0$ and collecting terms in Eq.~\ref{eqn_one} gives
\begin{widetext}
\begin{eqnarray}\label{detout_cal}
\mathcal{S}(\theta_{hwp},f) &=& \left(I + \frac{1}{2} Q \cos(4 \theta_{hwp})\right) T(f) \varepsilon(f)  + \left(\frac{1}{2} Q + \frac{1}{2} U \sin(4 \theta_{hwp})\right) T(f) \varepsilon(f) \gamma(f) \nonumber \\
 &+& \left(\frac{1}{2} Q \cos(2 \theta_{hwp}) + \frac{1}{2} U \sin(2 \theta_{hwp})\right) \rho(f) \varepsilon(f) + I \cos(2 \theta_{hwp}) \rho(f) \varepsilon(f) \gamma(f) \nonumber \\
 &+& \frac{1}{2}\left[Q  (1 -\cos(4 \theta_{hwp})) -  U \sin(4 \theta_{hwp}) \right] c(f) \varepsilon(f) \gamma(f),
\end{eqnarray}
\end{widetext}
where $I$, $Q$, and $U$ are the Stokes parameters of the light coming out of the FTS.  Placing a polarizing grid at the output of the FTS produces light with $I = 1$ and $Q = 1$. We call spectra taken in this configuration $\mathcal{S}_{0}$.  Rotating the grid by $45^{\circ}$ produces light with $I=1$ and $U=1$;  spectra taken with this configuration we call $\mathcal{S}_{45}$.

We have five parameters to constrain (for a single-plate system;  a multi-plate HWP has more), so we need at least five FTS spectra with different instrument and HWP angle combinations. Considering noiseless data, we need spectra with the polarizing grid at both 0 and $45^{\circ}$ to independently estimate all of the parameter combinations. With this data, we can set up a system of linear equations to estimate the HWP and detector parameters,
\begin{widetext}
\begin{equation}
\left[ \begin{array}{c} \mathcal{S}_{0}(\theta_{1},f)  \\  \mathcal{S}_{0}(\theta_{2},f) \\ \mathcal{S}_{0}(\theta_{3},f) \\ \vdots \\ \mathcal{S}_{45}(\theta_{N+1},f) \\ \mathcal{S}_{45}(\theta_{N+2},f) \\ \vdots \end{array} \right] = 
\frac{1}{2} \left[ \begin{array}{ccccc}
2+\cos(4 \theta_{1}) & 1 & \cos(2 \theta_{1}) & 2 \cos(2 \theta_{1}) & 1 - \cos(4 \theta_{1}) \\  
2+\cos(4 \theta_{2}) & 1 & \cos(2 \theta_{2}) & 2 \cos(2 \theta_{2}) & 1 - \cos(4 \theta_{2}) \\ 
2+\cos(4 \theta_{3}) & 1 & \cos(2 \theta_{3}) & 2 \cos(2 \theta_{3}) & 1 - \cos(4 \theta_{3}) \\ 
\vdots & \vdots & \vdots & \vdots & \vdots \\
2 & \sin(4 \theta_{N+1}) & \sin(2 \theta_{N+1}) & 2\cos(2 \theta_{N+1}) & - \sin(4 \theta_{N+1}) \\
2 & \sin(4 \theta_{N+2}) & \sin(2 \theta_{N+2}) & 2\cos(2 \theta_{N+2}) & - \sin(4 \theta_{N+2}) \\
\vdots & \vdots & \vdots & \vdots & \vdots \\
\end{array} \right] 
\left[ \begin{array}{c} T(f) \varepsilon(f) \textcolor{white}{\gamma(f)}\\ T(f) \varepsilon(f) \gamma(f) \\ \rho(f) \varepsilon(f)\textcolor{white}{\gamma(f)} \\ \rho(f) \varepsilon(f) \gamma(f) \\ c(f) \varepsilon(f) \gamma(f) \\ \end{array} \right] .
\end{equation}
\end{widetext}

Operationally, estimating this combination of the parameters is sufficient to characterize the full instrument 
(detectors + HWP) for polarized CMB mapmaking.  However, since the detector optical efficiency $\varepsilon(f)$ appears as an overall factor in these equations, we cannot independently estimate the HWP-only parameters $T(f)$, $\rho(f)$, and $c(f)$. To check the HWP physical optics model, we would need separate FTS spectra without the HWP in place to estimate the detector optical efficiency.

In this illustration, we do not consider possible parameter covariances or biases induced by detector noise and non-optimal instrument and HWP angles. One possible way to handle this would be to use FTS spectra taken at many HWP and instrument angles to obtain least-squares estimates of the calibration parameters and errors as a function of frequency. This would also allow a test in the $\chi^{2}$ sense on whether the performance of the HWP and instrument is well-described by our formalism.

\subsubsection{Rapidly-Rotating HWP}

If an instrument designed for a rapidly-rotating single-plate HWP has a mode in which the HWP can be stepped, then obviously the characterization procedure in the previous section may be used to estimate the HWP parameters.

As a substitute or complement, the HWP can be characterized in the instrument as it is rapidly rotating. 
For a single-plate HWP, taking $\psi_{inst} = 0^\circ$ and collecting the constant, $2 f_{hwp}$ and $4 f_{hwp}$ terms in Eq.~\ref{eqn_one} gives
\begin{eqnarray}\label{detout_psi_zero}
&&d(\theta_{hwp}) = \left[ T \varepsilon I + \frac{1}{2}(T+c) \varepsilon \gamma Q \right] \\
&+&  \left[ \frac{1}{2} (T - c) \varepsilon \gamma U \right] \sin{4 \theta_{hwp}} +  \left[ \frac{1}{2} (T - c) \varepsilon \gamma Q \right] \cos{4 \theta_{hwp}} \nonumber \\
&+& \left[ \rho \varepsilon \gamma I + \frac{1}{2} \rho \varepsilon Q \right] \cos{2 \theta_{hwp}} + \left[ s \varepsilon \gamma V + \frac{1}{2} \rho \varepsilon U \right] \sin{2 \theta_{hwp}} \nonumber.
\end{eqnarray}
This shows that only polarization information is contained in the $4f_{HWP}$ component of the detector timestream. A lock-in at $4 f_{HWP}$ on the timestream $d_{i}$ will produce companion $x_i$  and $y_i$ timestreams, where $x_{i}$ is the sine wave lock-in output timestream, and $y_{i}$ is the cosine timestream. The $Q$ and $U$ timestreams in instrument coordinates (since $\xi_{det} = \psi_{inst} = 0^\circ$) may be estimated using
\begin{eqnarray}
Q_i = \frac{2}{(T-c) \varepsilon \gamma} y_i \nonumber \\
U_i =\frac{2}{(T-c) \varepsilon \gamma} x_i 
\end{eqnarray}
This means that only one calibration factor is necessary to make polarization maps with a rapidly-rotating single-plate HWP. Doing this as a function of frequency with an FTS allows for the estimation of this calibration factor for different source spectral shapes.

\section{Polarized Mapmaking} \label{mapmaker}

In many CMB polarization experiments, the instrument continuously scans the sky, and the detector timestreams are used to estimate maps of the $I$, $Q$, and $U$ Stokes parameters.  Since the CMB is not expected to be circularly-polarized, we take $V=0$. To estimate the maps, a mapmaking algorithm such as the one in Jones et. al. \cite{jones08} assumes that without cross-polarization, the detector timestream $d_{i}$ can be modeled as
\begin{equation}\label{det_out_no_hwp}
d_i = \frac{1}{2} \left(I + Q \cos(2 \psi_i) + U \sin(2 \psi_i) \right) + n_i,
\end{equation}
where $\psi_{i}$ is the instrument angle timestream, and $n_i$ is atmospheric and detector noise. The algorithm then combines this detector timestream with the pointing timestream of the telescope, and uses an iterative matrix method to compute maximum likelihood maps of $I$, $Q$, and $U$, as well as an estimate of the noise timestream. The algorithm described in Jones et. al. \cite{jones08} also can treat cross polarization in the detector. 

As an illustration of the principle of the method applied to mapmaking with a HWP, we consider $N$ noiseless detector samples taken at different instrument angles $\psi_{i}$ and HWP angles $\theta_{i}$ when the telescope was pointed at the same spot on the sky. To estimate the Stokes parameters at that point on the sky, we set up a system of linear equations using the detector timestream model, 
\begin{equation}
\vec{d} = \mathsf{A} \left[ \begin{array}{c} I \\ Q \\ U \end{array} \right]
\end{equation}
Here $\vec{d}$ is a $N$-element column vector of detector samples, and $\mathsf{A}$ is a $N$-by-$3$ matrix containing the detector and HWP model.
If we take Eq.~\ref{det_out_no_hwp} and model an ideal polarized detector with an ideal HWP, the $\mathsf{A}$ matrix is
\begin{equation}
\mathsf{A}_{ideal} = \frac{1}{2} \left[ \begin{array}{ccc} 1 & \cos{(2 (\psi_1 + 2 \theta_1))} & \sin{(2 (\psi_1 + 2 \theta_1))} \\ 1 & \cos{(2 (\psi_2 + 2 \theta_2))} & \sin{(2 (\psi_2 + 2 \theta_2))} \\ \vdots & \vdots & \vdots \\ 1 & \cos{(2 (\psi_N + 2 \theta_N))} & \sin{(2 (\psi_N + 2 \theta_N))} \\  \end{array} \right].
\end{equation}
To estimate the polarization state of the incident light, we invert the $\mathsf{A}$ matrix,
\begin{equation}\label{pol_est}
\left[ \begin{array}{c} I \\ Q \\ U \end{array} \right] = \mathsf{A}^{-1} \vec{d}.
\end{equation}

To treat the effects of the HWP non-idealities, we can use the timestream model of Eq.~\ref{eqn_one} and the Mueller matrix elements from Eq.~\ref{mat_multiply} to set up the relevant system of linear equations. In this case the $\mathsf{A}$ matrix becomes
\begin{equation}
\mathsf{A}_{real} = \left[ \begin{array}{ccc} M_{II}(\theta_1,\psi_1) & M_{IQ}(\theta_1,\psi_1) & M_{IU}(\theta_1,\psi_1) \\ M_{II}(\theta_2,\psi_2) & M_{IQ}(\theta_2,\psi_2) & M_{IU}(\theta_2,\psi_2) \\ \vdots & \vdots & \vdots \\ M_{II}(\theta_N,\psi_N) & M_{IQ}(\theta_N,\psi_N) & M_{IU}(\theta_N,\psi_N) \\ \end{array} \right].
\end{equation}
Even though this matrix has more complicated elements, Eq.~\ref{pol_est} still holds and we can still estimate the polarization state of the incident light with a matrix inversion. Thus, simply changing the functional form of the 
$\mathsf{A}$ matrix in the mapmaking algorithm allows it to exactly compensate for the HWP non-idealities in our model.

\section{Application to the SPIDER HWP}

\subsection{Mueller Matrix}
The HWPs in the SPIDER instrument will consist of a single 330 mm diameter single-crystal sapphire plate for each telescope with a quarter-wave quartz layer attached to each sides. Each HWP will be cooled to $\sim 4$ K. Here we consider the prototype HWPs for the SPIDER 145 GHz band. 

We have measured the indices of refraction of sapphire at 5 K to be $n_s~=~3.336~\pm~.003$ and $n_f~=~3.019~\pm~.003$, from 100 GHz to 240 GHz~\cite{bryan10a}.  Our fused quartz AR coats  have a room temperature index of refraction of $n_{ar} = 1.951$ at 245~GHz~\cite{lamb95}.  We use Eq.~\ref{hwp_jones_calculation} to calculate the Mueller matrix of a thickness-optimized SPIDER 145 GHz HWP as a function of frequency; the results are shown in Fig.~\ref{hwp_mueller}.  We assumed all materials were lossless, though that assumption can easily be relaxed and the output used in the Mueller formalism described above without modification.


We then use Eq.~\ref{band_average} to calculate the band-averaged matrix elements of the SPIDER HWP, 
assuming a top-hat detector spectrum from 130 GHz to 160 GHz. For the source spectra we use
\begin{align}
S(f) \propto 
\begin{cases} 
  1 & \mathrm{Flat} \\
  \frac{dB}{dT} (f,2.725~\mathrm{K})  & \textrm{CMB~\cite{fixsen03}} \\
  f^{1.67} B(f,9.6~\mathrm{K}) \\
  ~~~+ 0.0935 f^{2.7} B(f,16.2~\mathrm{K}) & \textrm{Dust~\cite{finkbeiner99}} \\
  f^{-1} & \textrm{Synchrotron~\cite{bennett03}} \\
  f^{-.14} & \textrm{Free-free~\cite{oster61}},
\end{cases}
\end{align}
where $B(f,T)$ is the blackbody function, 
as estimates of the CMB and astrophysical foregrounds.  The results are shown in Table \ref{hwp_params}.  The $c$ parameter deviates from ideality by almost $5\%$, which is a relatively large effect. This pushes us towards handling the non-idealities through calibration and a modified mapmaker as discussed in Section \ref{mapmaker}.

\begin{table}
\begin{center}
\begin{tabular}{ c | c | c | c | c|}
 & $T$ & $\mathbf{\rho}$ & $c$  & $s$\\
\hline
\hline
\textbf{Flat} & 0.97389 & 0.01069 & -0.95578 & \phantom{-}0.00170 \\
\hline
\textbf{CMB} & 0.97396& 0.01069 & -0.95591 & -0.00952 \\
\hline
\textbf{Dust} & 0.97391 & 0.01080 & -0.95563 & -0.03598 \\
\hline
\textbf{Synchrotron} & 0.97382 & 0.01070 & -0.95565 & \phantom{-}0.01280\\
\hline
\textbf{Free-free} & 0.97388 & 0.01069 & -0.95577 & \phantom{-}0.00325 \\
\hline
\textrm{(Ideal HWP)} & 1 & 0 & -1 & 0\\
\hline
\end{tabular}
\caption{Calculated Mueller matrix elements for an optimized cryogenic sapphire HWP with a quartz AR-coat for the \textit{Spider} 145 GHz HWP. We optimized the HWP thickness based on our measured cold indices, and the AR-coat thickness based on the index in Lamb~\cite{lamb95}. The first row shows the HWP parameters averaged within the \textit{Spider} 145 GHz passband. The CMB, Dust, Synchrotron, and Free-free rows all are band-averaged against the source spectra within the passband. The last row shows the parameter values of an ideal HWP for comparison. \label{hwp_params}}
\end{center}
\end{table}

In addition to being large amplitude, the $s$ parameter depends significantly on which source spectrum is used;  this is not surprising given the form of $s$ across the band, shown in Fig.~\ref{hwp_params}, which shows that tailoring $s$ to be near zero requires near symmetric placement and weighting across the band, since $s$ is large but asymmetric.
The variations of $T$, $\rho$, and $c$ between the CMB and foreground sources are not as worrisome.  The parameters vary among the different sources at less than the $10^{-4}$ level. 

For comparison, SPIDER is targeting $B$-modes at the $r\sim.03$ level, and we will remove dust foregrounds by combining maps from several frequencies. To reach our target $B$-mode sensitivity, our goal for dust foreground subtraction is to reduce it by $\sim 95\%$. The HWP response varies among the different sources at the $10^{-4}$ level, which suggests that residuals from foreground subtraction should be dominated by incomplete dust removal, not source-varying HWP systematics. Next-generation experiments targeting $r<.01$ will require better foreground removal, and may be impacted by this effect.

\subsection{Mapmaking}
\begin{figure*}
\begin{center}
\includegraphics[width=\textwidth]{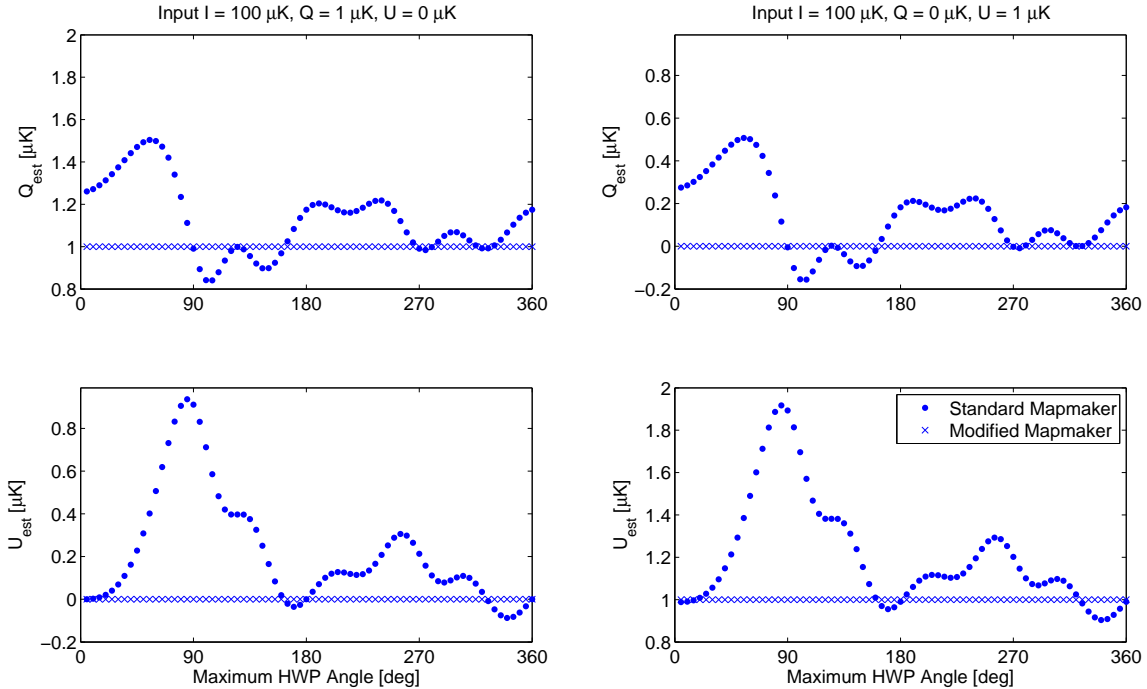}
\caption{$Q$ and $U$ Stokes parameter estimates made with a conventional mapmaker and a mapmaker modified to handle the SPIDER HWP non-idealities. Each estimate was made with simulated detector samples from 10 HWP angles evenly spaced from $0^\circ$ to $\theta_{max}$, where $\theta_{max}$ is the horizontal axis of each plot. The conventional mapmaker, plotted in dots, misestimates the $Q$ and $U$ Stokes parameters by up to $1~\mu$K. The main reason for this is the temperature-polarization leakage in the HWP. The modified mapmaker, plotted as crosses, completely handles the non-idealities modeled in our formalism. \label{hwp_mapmaker_comparison}}
\end{center}
\clearpage 
\end{figure*}

A full time-domain simulation of a given instrument's scan strategy is necessary to precisely quantify the effects of using a conventional mapmaking algorithm with a non-ideal HWP, and to see if the modified mapmaker described above improves the results. To get a sense for what the results might be, we performed a simpler simulation. We input a Stokes vector of $[100~\mu \mathrm{K},1~\mu \mathrm{K},0,0]$ to the detector model in Eq.~\ref{eqn_one}. This Stokes vector is representative of the amplitudes of temperature and $E$-mode CMB fluctuations at the angular scales relevant for SPIDER. We fixed $\xi_{det} = \psi_{inst} = 0^\circ$ and generated one detector sample at each of the HWP angles $0^\circ, 1\times \frac{\theta_{max}}{10-1}, 2\times \frac{\theta_{max}}{10-1}, \dots, \theta_{max}$ to make a total of 10 samples evenly spaced in angle from $0^{\circ}$ to $\theta_{max}$. We used these simulated detector samples to estimate the $Q$ and $U$ Stokes parameters using Eq.~\ref{pol_est}. We simulated the effect of using a conventional mapmaker by using the matrix $A_{ideal}$, and simulated a modified mapmaker by using $A_{real}$.  We repeated this process with varying $\theta_{max}$ to generate Fig.~\ref{hwp_mapmaker_comparison}, which is a plot of reconstructed $Q$ and $U$ as a function of $\theta_{max}$. We generated the second pair of plots shown in Fig.~\ref{hwp_mapmaker_comparison} with an input Stokes vector of $[100~\mu \mathrm{K},0,1~\mu \mathrm{K},0]$.

Not shown in Fig.~\ref{hwp_mapmaker_comparison} is the fact that the conventional mapmaker 
also mis-estimates the $I$ Stokes parameter when using samples from a detector with a non-ideal HWP. 
However, most CMB instruments (including SPIDER) will calibrate by cross-correlating the intensity maps with the WMAP~\cite{jarosik10} or upcoming Planck satellite results, which would change the overall calibration to compensate for the $I$ mis-estimation.  To approximate the effect of this calibration process, we scaled our estimates of $I$, $Q$ and $U$ made with the conventional mapmaker by a factor to force the $I$ estimate to be the same as the input value. This was not necessary for the modified mapmaker since it accurately estimates the Stokes $I$ term. 
The reason the conventional mapmaker generates the large residuals shown in Fig.~\ref{hwp_mapmaker_comparison} is that it does not handle the temperature-polarization leakage generated by the HWP.

Quantifying the systematics from using a non-ideal HWP with a conventional mapmaker will require full time-domain instrument simulations. These will be presented for the case of SPIDER in an upcoming paper 
\cite{odea_future}. 


\section{Conclusions}
We have derived an analytic model for birefringent, AR-coated HWPs that simulates the response of an instrument to different source spectra on the sky at arbitrary instrument and HWP angles. We also modeled the sum and difference timestreams of a PSB pair looking through a HWP, and showed that it should be possible to characterize the HWP and detectors with lab testing of a completed instrument.  Motivated by the presence of potentially significant non-idealities in the HWP, we presented a mapmaking algorithm that accounts for these known  non-idealities of the HWP.   This model will be integrated into an end-to-end simulation of the SPIDER instrument to translate these non-idealities into detailed limits on their possible contamination of power spectral and parameter estimates.


After the initial calculation of band-averaged parameters, our Mueller matrix method is analytic and
does not require repetitive matrix multiplications or repetitive integration over frequency.  Therefore, 
modeling detector timestreams is far faster than repeated use of a Jones-formalism code. 
As an example, in 10 seconds of computer time in Matlab on a laptop, the direct Jones method can simulate 300 SPIDER detector samples, while the Mueller matrix method can simulate 24,000, a speedup by a factor of $\sim75$. A version of the Mueller matrix code written in C for use in the instrument simulation code for SPIDER can simulate 10 million detector samples in 10 seconds on the same machine.

\section{Acknowledgements}
The authors are supported by NASA grant number NNX07AL64G. We would like to thank the anonymous referee for the helpful suggestions.


\clearpage

\end{document}